\documentclass[prb,twocolumn, showpacs]{revtex4}
\usepackage{amsmath}
\usepackage{graphicx}
\def\ket#1{\left|#1\right\rangle}
\def\bra#1{\left\langle#1\right|}

\begin{document}

\title{Quantum Theory of a Resonant Photonic Crystal}

\author{Y.~D.~Chong}
\email{cyd@mit.edu}
\author{David E.~Pritchard}
\author{Marin Solja\v{c}i\'{c}}

\affiliation{Department of Physics, Massachusetts Institute of
  Technology, Cambridge, Massachusetts 02139}

\date{\today}

\begin{abstract}
We present a quantum model of two-level atoms localized in a
3\textsc{D} lattice, based on the Hopfield polariton theory.  In
addition to a polaritonic gap at the excitation energy, a photonic
bandgap opens up at the Brillouin zone boundary.  Upon tuning the
lattice period or angle of incidence to match the photonic gap with
the excitation energy, one obtains a combined polaritonic and photonic
gap as a generalization of Rabi splitting.  For typical experimental
parameters, the size of the combined gap is on the order of
$25\,\textrm{cm}^{-1}$, up to $10^5$ times the detuned gap size.  The
dispersion curve contains a branch supporting slow-light modes with
vanishing probability density of atomic excitations.
\end{abstract}


\maketitle

\section{Introduction}

Photonic crystals (PhCs)–--systems in which the index of refraction
varies periodically on the scale of light---are known to have an
extraordinary ability to control the flow of light
\cite{Sajeev,Yablonovitch,JJ}.  While the underlying index of
refraction in conventional PhCs is often taken to be the bulk value,
interesting effects can occur when the underlying medium possesses
resonances at wavelengths comparable to the lattice spacing; for
instance, one can dramatically widen the photonic bandgap by tuning
the bandgap frequency (e.g.~by changing the lattice period) to match
the resonance frequency.  Such ``resonant PhCs'' can be realized using
cold atoms in optical lattices \cite{Jessen,Deutsch,Coevorden}, PhCs
made from polaritonic materials \cite{Casey}, and
multiple-quantum-well arrays \cite{MQW,MQW2,MQW3}.  Here, we will
concentrate on the first class of resonant PhCs, originally analyzed
by Deutsch, Spreeuw, Rolston, and Phillips\cite{Deutsch}, who modelled
the atoms in a 1\textsc{D} optical lattice as a set of classical
polarizable planes and showed that the interaction of the resonances
with the periodicity of the system gives rise to a photonic bandgap.
Subsequently, van Coevorden \textit{et.~al.}\cite{Coevorden}~extended
this study to 3\textsc{D} by solving Maxwell's equations in a lattice
of resonating point dipoles using a t-matrix analysis.

In this paper, we present a simple 3\textsc{D} quantum mechanical
model of an atomic PhC in which the elementary excitations are
polaritons: coherent superpositions of atomic excitations and photons.
Several features of previous classical models appear naturally, and
with somewhat simpler interpretations, in the quantum model.  For
instance, we show that the resonance-induced bandgap arises as a
generalization of Rabi splitting in a microcavity.  Our model also
exhibits the important ``Bragg resonant'' modes first studied in
1\textsc{D} by Deutsch \textit{et.~al.}, who identified them with the
standing electromagnetic wave that supports the optical
lattice\cite{Deutsch}.  Here, the Bragg resonant modes generalize to a
family of modes occupying the boundary of the first Brillouin zone
(BZ), and attached to the dispersion curve associated with atomic
excitations; near the BZ boundary, they possess low group velocity but
involve little excitation of the underlying medium, unlike states in
``slow light'' systems\cite{slow light}.

\section{Model Hamiltonian}

Consider $N$ localized two-level atoms of the same type in a
fully-filled 3\textsc{d} cubic lattice, at sites $\vec{r}_i$ with
lattice period $\ell$.  To facilitate calculation, we enclose the
lattice in a periodic electromagnetic cavity of volume $V$, which
reproduces the physical behavior inside a sufficiently large lattice.
The Coulomb-gauge Hamiltonian is
\begin{equation}
  H = \sum_i \epsilon \, b^\dagger_i b_i + \sum_{\vec{k}\sigma} \hbar c
  |\vec{k}| \, a^\dagger_{\vec{k}\sigma} a_{\vec{k}\sigma} - \frac{e}{mc} \sum_i
  \vec{A}(\vec{r}_i) \cdot \vec{p}_i,
\label{QED Hamiltonian}
\end{equation}
where $\epsilon$ is the energy difference between the atomic levels,
$b^\dagger_i\!\equiv\!\ket{1}\!\bra{0}_i$ and $b_i\!\equiv\!
\ket{0}\!\bra{1}_i$ are the level raising and lowering operators for
atom $i$, and $a^\dagger_{\vec{k}\sigma}$ and $a_{\vec{k}\sigma}$ are
creation and annihilation operators for photons with wavevector
$\vec{k}$ and polarization $\sigma$.  $\vec{A}(\vec{r})$ is the vector
potential,
\begin{equation}
  \vec{A}(\vec{r}) = \sum_{\vec{k}\sigma}
  \sqrt{\frac{2\pi\hbar c}{V|\vec{k}|}}\;
  \left( a_{\vec{k}\sigma}\,e^{i \vec{k}\cdot \vec{r}}
            + a^\dagger_{\vec{k}\sigma}\,e^{-i \vec{k}\cdot \vec{r}}
            \right) \hat{e}_{\vec{k}\sigma},
  \label{A field}
\end{equation}
where $\hat{e}_{\vec{k}\sigma}$ is the unit polarization vector for
$a^\dagger_{\vec{k}\sigma}$.

Let us suppose that the average number of atomic excitations in the
system at any time is much less than $N$.  In that case, the atomic
excitations are approximately bosonic, in the same sense that spin
waves are bosons\cite{Hopfield}.  Therefore, the two photon
polarizations, which excite orthogonal atomic states, decouple for
each $k$.  We thus drop the $\sigma$ label, with the understanding
that the dispersion relations we will later obtain are doubly
degenerate.  This is also consistent with the weak polarization
dependence obtained by van Coevorden \textit{et.~al.}\cite{Coevorden}
In contrast, polarization effects play an important role in
conventional PhCs\cite{JJ}, as well as multiple-quantum-well resonant
PhCs\cite{MQW,MQW2,MQW3}, due to the finite size of the scattering
centers.

We can treat the $\vec{r}_i$ in (\ref{QED Hamiltonian}) as numbers
(perfect lattice positions) rather than operators, since the
electronic wavefunctions are typically much narrower than the lattice
spacing.  At each site, the momentum operator is
\begin{equation}
  \vec{p}_i = \frac{i}{\hbar}\, m \epsilon \, \vec{x}_{01} (b^\dagger_i - b_i)\;,\quad
  \vec{x}_{01} \equiv \langle1|\vec{x}|0\rangle.
  \label{site momentum}
\end{equation}
Let us also define momentum-space excitation operators
\begin{equation}
  b_{\vec{q}} = \frac{1}{\sqrt{N}} \sum_i e^{-i \vec{q} \cdot \vec{r}_i}\,b_i \;,\;
  b^\dagger_{\vec{q}} = \frac{1}{\sqrt{N}} \sum_i e^{i \vec{q} \cdot \vec{r}_i}\,b^\dagger_i,
  \label{b operators}
\end{equation}
where the wavevectors $\vec{q}$ are restricted to the first BZ,
corresponding to the fact that an excitation ``wave'' has no meaning
between lattice points.  As explained above, these are approximately
bosonic: $[b_{\vec{q}},b_{\vec{q}'}^\dagger] \simeq
\delta_{\vec{q}\vec{q}'}$.

Substituting (\ref{A field})-(\ref{b operators}) into (\ref{QED
  Hamiltonian}), we obtain the microscopic polariton Hamiltonian first
derived by Hopfield \cite{Hopfield} in the context of crystalline
solids:
\begin{multline}
  H = \sum_{\vec{q}} \biggl\{ \epsilon \, b^\dagger_{\vec{q}} \, b_{\vec{q}} +
          \sum_{\vec{\textsc{g}}} \hbar c \,|\vec{q}+\vec{\textsc{g}}|\,
          a^\dagger_{\vec{q}+\vec{\textsc{g}}}\,a_{\vec{q}+\vec{\textsc{g}}} \\
          - \sum_{\vec{\textsc{g}}} i\, C_{\vec{q}+\vec{\textsc{g}}} \left[
            \left(b^\dagger_{\vec{q}} \, a_{\vec{q}+\vec{\textsc{g}}} -
            a^\dagger_{\vec{q}+\vec{\textsc{g}}} \, b_{\vec{q}} \right) \right.\\
            \left.+ 
            \left(b^\dagger_{\vec{q}} \, a^\dagger_{-(\vec{q}+\vec{\textsc{g}})}
            - a_{-(\vec{q}+\vec{\textsc{g}})} \, b_{\vec{q}} \right) \right]
          \biggl\},
\label{Hopfield Hamiltonian}
\end{multline}
where $\vec{\textsc{g}}$'s run over all reciprocal lattice vectors,
and
\begin{equation}
  C_{\vec{q}+\vec{\textsc{g}}} = \sqrt{\frac{2\pi \alpha N}{|\vec{q}+\vec{\textsc{g}}|\,V}}
  \; \epsilon \,x_{01}
\label{coupling strength}
\end{equation}
where $\alpha$ is the fine structure constant.  The atom-photon
interaction consists of two parts.  The first part, on the second line
of (\ref{Hopfield Hamiltonian}), describes the lattice absorbing a
photon with wavevector $\vec{q}+\vec{\textsc{g}}$ to create an atomic
excitation with wavevector $\vec{q}$, and the reverse process of
destroying an excitation to emit a photon.  The remaining interaction
terms describe the creation and annihilation of associated pairs of
photons and atomic excitations.  The usual way to diagonalize
(\ref{Hopfield Hamiltonian}) is to introduce polariton operators
$\alpha_{\vec{q}}$ \cite{Hopfield, KandM} for each reduced wavevector
$\vec{q}$, as linear combinations of $b_{\vec{q}}$,
$b^\dagger_{-\vec{q}}$, $a_{\vec{q}+\vec{\textsc{g}}}$, and
$a^\dagger_{-\vec{q}+\vec{\textsc{g}}}$ (for all $\vec{\textsc{g}}$).
Stipulating that these act as decoupled lowering operators for $H$,
one obtains the polariton energies as solutions of a
$(2n+1)\times(2n+1)$ eigenvalue problem for each $\vec{q}$, where $n$
is the number of BZs included in the calculation.  Higher BZs were
first included into the Hopfield theory by Knoester and Mukamel
\cite{KandM} in their calculation of polariton-mediated intermolecular
forces in solids.  There, the photons in the higher BZs were taken to
be decoupled from the atomic excitations, which was appropriate since
the BZ energy was many orders of magnitude larger than $\epsilon$.  In
our system, the two energies are comparable, and we must incorporate
the interaction up to at least the second-order zones.

It simplifies the calculations to drop the ``counter-rotating''
interaction terms in (\ref{Hopfield Hamiltonian}) describing the
creation and annihilation of pairs.  This is physically justifiable
even though the discarded terms have the same coupling strength
$C_{\vec{q}+\vec{\textsc{g}}}$ as the remaining interaction terms,
because the pair creation and annihilation process is a quantum
mechanical fluctuation of the ``vacuum'' with a finite energy gap
$\epsilon + \hbar c|\vec{q}|$.  For $\epsilon$ and $\hbar c |\vec{q}|$
both on the order of eV, and lattice periods at optical wavelengths,
$C_{\vec{q}} \sim 10^{-4}$\,eV $\ll \epsilon + \hbar c|\vec{q}|$.
Such fluctuations are thus extremely rare and have a negligible effect
on particle energies.  The interaction terms describing the conversion
of a real photon into an atomic excitation, and vice versa, remain
important: since the existing particle possesses energy, these
processes involve a much smaller energy fluctuation.  The
approximation holds provided we look at values of $|\vec{q}|$
comparable to both $\epsilon/\hbar c$ and the BZ energy, which is
exactly the regime we are interested in.

\begin{figure}
\includegraphics[width=0.48\textwidth]{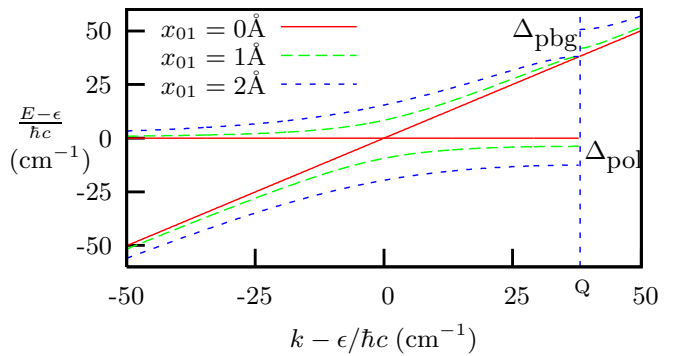}
\caption{(Color online) Single-polariton dispersion for a 3\textsc{d} cubic lattice
  along $[1 0 0]$ in the extended zone scheme, with $\epsilon = 3$\,eV
  and three different coupling strengths, associated with the
  parameters $x_{01} = 0$ (non-interacting), 1\,\AA\,($C_{\textsc{q}}
  \sim 0.18$\,meV), and 2\,\AA\,($C_{\textsc{q}} \sim 0.35$\,meV).
  The vertical dashed line indicates the BZ boundary at
  $|\vec{\textsc{q}}| = 1.00025\,\epsilon/\hbar c$.  The graphs are
  generated numerically from (\ref{our Hamiltonian}), summing over 125
  BZs.  }
\label{polariton interaction figure}
\end{figure}

The Hamiltonian now decouples into $N$ independent pieces, $H
=\sum_{\vec{q}} H_{\vec{q}}$, one for each reduced wavevector:
\begin{align}
\begin{aligned}
  H_{\vec{q}} =& \;\epsilon \, b^\dagger_{\vec{q}} \, b_{\vec{q}} +
  \sum_{\vec{\textsc{g}}} \hbar c \,|\vec{q}+\vec{\textsc{g}}|\,
  a^\dagger_{\vec{q}+\vec{\textsc{g}}} \, a_{\vec{q}+\vec{\textsc{g}}} \\
  &- \sum_{\vec{\textsc{g}}}
  i\, C_{\vec{q}+\vec{\textsc{g}}} \left(b^\dagger_{\vec{q}} \,
  a_{\vec{q}+\vec{\textsc{g}}} - a^\dagger_{\vec{q}+\vec{\textsc{g}}} \, b_{\vec{q}}
  \right).
  \label{our Hamiltonian}
\end{aligned}
\end{align}
This says that each photon mixes with all other photons having
wavevectors that differ by a reciprocal lattice vector, as one expects
of a PhC system.  Here, the mixing is mediated by the atom-photon
interaction.  Since (\ref{our Hamiltonian}) has the quadratic form
$\sum_{ij} \beta_i^\dagger \mathcal{H}_{ij} \beta_j$, it can be
diagonalized as $\sum_n E_n \alpha_n^\dagger \alpha_n$, where the
$\alpha$'s are boson operators defined by $\alpha_n = \sum_j w_j^{*n}
\beta_j$, $E_n$ is the $n$th eigenvalue of $\mathcal{H}$, and $w^n$ is
the corresponding eigenvector.  We can thus obtain the polariton
energies $E^n_{\vec{q}}$ by including a finite number of BZs in the
sum and diagonalizing the associated matrix.

\section{Band structure}

\begin{figure}
\includegraphics[width=0.48\textwidth]{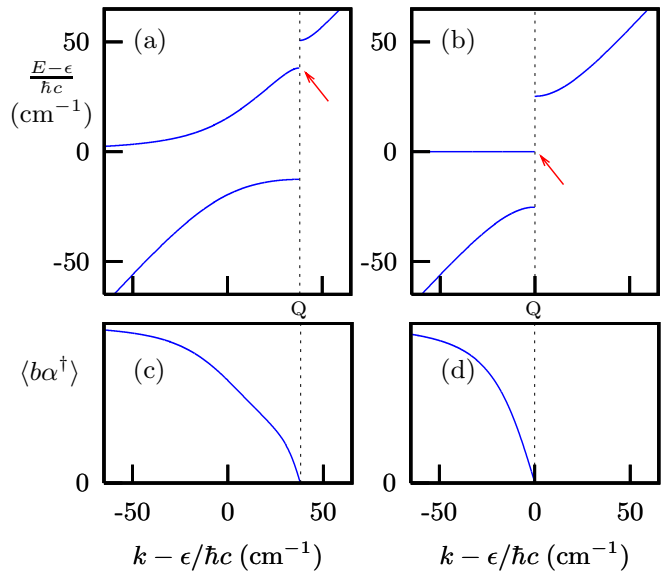}
\caption{(Color online) Single-polariton dispersion along $[1 0 0]$,
  with $\epsilon = 3$\,eV, $x_{01} = 2$\,\AA, and different lattice
  periods: (a) $|\vec{\textsc{q}}| = 1.00025\,\epsilon/\hbar c$, and
  (b) $|\vec{\textsc{q}}| = \epsilon/\hbar c$.  Plots (c) and (d) show
  the corresponding overlaps of the polariton with the bare
  excitation, $\bra{0}b_{\vec{q}}\alpha^\dagger_{\vec{q}}\ket{0}$, for
  the polaritons on the dispersion curve leading to the purely
  photonic state at $\vec{q} = \vec{\textsc{q}}$ (indicated with
  arrows in (a) and (b)), which have no atomic component.}
\label{polariton figure}
\end{figure}

Fig.~\ref{polariton interaction figure} shows the polariton dispersion
curves along the $[1 0 0]$ direction for a blue-detuned optical
lattice.  The interaction opens up two energy gaps in the polariton
spectrum: an indirect ``polaritonic gap'' $\Delta_{\textrm{pol}}$ at
$\epsilon$ due to the repulsion between the bare dispersion curves,
and a photonic bandgap $\Delta_{\textrm{pbg}}$ at $\hbar c
|\vec{\textsc{q}}|$ where $\vec{\textsc{q}}$ is the BZ boundary.  We
have also calculated the density of polariton states; after
integrating over all angles, we find that the density of states is
enhanced near the band edges, but remains nonzero at all energies
because the exact sizes and positions of the gaps vary with angle.
The system therefore does not possess a complete gap, essentially
because of the weakness of the electromagnetic interaction.  The gap
sizes vary continuously as we change the lattice period $a$, and thus
$V$ (keeping $N$ and all other parameters constant).  As shown in
Fig.~\ref{polariton figure}, the gaps meet and become significantly
enhanced when the BZ boundary intersects the crossing point of the
bare dispersion curves.

To understand the nature of the spectrum at the BZ boundary, consider
a photon with wavevector $\vec{k} = \vec{\textsc{q}}$ along one of the
faces of the cube.  There is another such photon, with wavevector
$\vec{\textsc{q}}+\vec{\textsc{g}}'$ lying on the opposite face, such
that $|\vec{\textsc{q}}| = |\vec{\textsc{q}}+\vec{\textsc{g}}'|$.
(When $\vec{\textsc{q}}$ lies on an edge or corner of the BZ boundary,
there are more partners; we will not consider these cases, but they
can be treated in a similar fashion.)  The two photons mix strongly
since they have the same energy, so we can neglect the other photon
states and use the effective Hamiltonian
\begin{equation}
  \tilde{H}_{\vec{\textsc{q}}} =
  \begin{bmatrix} b_{\vec{\textsc{q}}} \\ a_{\vec{\textsc{q}}}
    \\ a_{\vec{\textsc{q}}+\vec{\textsc{g}}'}
  \end{bmatrix}^\dagger
  \begin{bmatrix}
    \epsilon & -iC_{\vec{\textsc{q}}} & -iC_{\vec{\textsc{q}}} \\
    iC_{\vec{\textsc{q}}} & \hbar c|\vec{\textsc{q}}| & 0 \\
    iC_{\vec{\textsc{q}}} & 0 & \hbar c|\vec{\textsc{q}}|\\
  \end{bmatrix}
  \begin{bmatrix} b_{\vec{\textsc{q}}} \\ a_{\vec{\textsc{q}}} \\
    a_{\vec{\textsc{q}}+\vec{\textsc{g}}'}
  \end{bmatrix}.
  \label{edge Hamiltonian}
\end{equation}
Thus, the polariton energies at the BZ boundary are
\begin{align}
\begin{aligned}
  E^0_{\vec{\textsc{q}}} &= \hbar c |\vec{\textsc{q}}|, \\
  E^{\pm}_{\vec{\textsc{q}}} &= \frac{\epsilon+\hbar c |\vec{\textsc{q}}|}{2} \pm
             \sqrt{\left(\frac{\epsilon - \hbar c |\vec{\textsc{q}}|}{2}\right)^2
                    + 2 C_{\vec{\textsc{q}}}^2}\;.
\end{aligned}
  \label{edge energies}
\end{align}
These are exactly the energy levels resulting from Rabi splitting of a
two-level atom interacting with two counterpropagating photon states
with wavevectors $\pm\vec{\textsc{q}}$, with an effective cavity size
$V/N$.  In the exactly-tuned case $\epsilon = \hbar c
|\vec{\textsc{q}}|$, $E^\pm_{\textsc{q}}$ has a special significance:
as shown in Fig.~\ref{polariton figure}(b), these are the upper and
lower edges of the bandgap.  The resonant enhancement of the bandgap
in this system is thus a manifestation of the Purcell effect
\cite{Purcell}.  Intuitively, we can imagine enclosing a single atom
in a microcavity with the dimensions of the unit cell; if the cavity
walls are mirrors, the atom sees a lattice of atoms similar to the one
considered here.

We have checked (\ref{edge energies}) against numerical
solutions of (\ref{our Hamiltonian}) including the 125 lowest BZs, for
various values of $\vec{\textsc{q}}$ along the BZ boundary up to
$40^\circ$ from the $[1 0 0]$ direction.  For $\epsilon = 3$\,eV and
$x_{01} = 2$\,\AA, the error is always less than
$0.02\,\textrm{cm}^{-1}$, three orders of magnitude smaller than the
maximum gap size.

The size of the gaps in the exactly-tuned limit can be estimated by
substituting $\epsilon = \hbar c |\vec{\textsc{q}}'|$ into (\ref{edge
  energies}):
\begin{equation}
  \Delta \approx \sqrt{2} \; C'_{\epsilon/\hbar c}
  = \sqrt{\frac{4\alpha x_{01}^2 \epsilon^4}{\pi^2(\hbar c)^2}}.
  \label{large gap}
\end{equation}
For $\epsilon, \hbar c |\vec{\textsc{q}}| \approx 3$\,eV, and $x_{01}
\approx 2$\,\AA, $\Delta/\hbar c \approx 25\,\textrm{cm}^{-1}$ ($\sim
10^{-4} \, \epsilon$), in agreement with Fig.~\ref{polariton
  figure}(b).  We can also obtain limiting expressions for the gaps
when they are significantly decoupled.  Consider $|\vec{\textsc{q}}| >
\epsilon/\hbar c$, as in Fig.~\ref{polariton figure}(a).  Away from
the BZ boundary, we can neglect the effect of photons in higher BZs,
and the effective Hamiltonian matrix is $\mathcal{H} =
[\,\epsilon,\,-iC_{\vec{q}}\,;\,iC_{\vec{q}},\,\hbar c|\vec{q}|\,]$,
with eigenvalues
\begin{equation}
  E^{\pm}_{\vec{q}} = \frac{\epsilon+\hbar c |\vec{q}|}{2} \pm
             \sqrt{\left(\frac{\epsilon - \hbar c |\vec{q}|}{2}\right)^2
                    + C_{\vec{q}}^2}\;.
  \label{polariton dispersion}
\end{equation}
The contribution to the indirect polaritonic gap from the large-$q$
branch of the dispersion curve, which is truncated at the BZ boundary,
is obtained from the large-$q$ expansion of (\ref{polariton
  dispersion}) evaluated at $\vec{q} = \vec{\textsc{q}}$.  The
contribution from the small-$q$ branch cannot be found by setting
$\vec{q} = 0$ in (\ref{polariton dispersion}) due to our preceding
approximations, so we instead calculate an upper bound on it by
evaluating it at the minimum, $|\vec{q}| = \epsilon / 2\hbar c$.  The
resulting polaritonic gap is
\begin{equation}
  \Delta_{\textrm{pol}}' \simeq \frac{4 C^2_{\epsilon/\hbar c}}{\epsilon}
                + \frac{C^2_{\vec{\textsc{q}}}}{\hbar c \textsc{q}}
\label{small gap}
\end{equation}
With the same lattice parameters, $\Delta'_{\textrm{pol}} \approx
10^{-3}\,\textrm{cm}^{-1}$ ($\sim 10^{-8} \, \epsilon$).  From the
large-$|\vec{\textsc{q}}|$ expansion of (\ref{edge energies}), the
photonic bandgap is $\Delta_{\textrm{pbg}}' =
C^2_{\vec{\textsc{q}}}/\hbar c |\vec{\textsc{q}}|$, strictly smaller
than (\ref{small gap}).  Therefore, the effects of the polaritonic
interaction are very small when the system is detuned.

This model can also be used to study the quasi-1D geometry considered
by many authors, in which atoms are trapped along periodically-stacked
infinite sheets.  Consider a 3\textsc{d} lattice in which the lattice
spacing in one of the directions, $\ell_1$, is much larger than the
spacing in the other two directions.  The relevant wavevectors, lying
on the BZ boundaries closest to the origin, have magnitude
$|\vec{\textsc{q}}_1| = \pi/\ell_1$ and point in the direction of
stacking.  In this regime, this model can be directly compared with
the semiclassical analysis of Deutsch \textit{et.~al.}\cite{Deutsch}.
For instance, the semiclassical theory predicts bandgaps from
$E_{-}^{(cl)}$ to $\epsilon$ and from $\hbar c |\vec{\textsc{q}}_1|$
to $E_{+}^{(cl)}$ for blue-detuned lattices.  A short calculation,
using Eq.~15-19 of that paper, yields
\begin{equation}
  E_{\pm}^{(cl)} \approx \frac{\epsilon+\hbar c |\vec{\textsc{q}}|}{2} \pm
  \sqrt{\left(\frac{\epsilon - \hbar c |\vec{\textsc{q}}|}{2}\right)^2
    + 2 \cdot \frac{3\hbar^2c\gamma\eta}{2|\vec{\textsc{q}}_1|}}
\end{equation}
where $\eta$ is the surface density along each sheet and $\gamma \ll
(E_{\pm} - \epsilon)/\hbar$ is the linewidth of the atomic transition.
Using the golden rule prescription for the natural
linewidth\cite{Fermi}, $\gamma = (4\alpha \epsilon^3
x_{01}^2)/(3\hbar^3 c^2)$, this reduces to (\ref{edge energies}) with
$C_{\textsc{q}}^2$ replaced by $C_{\textsc{q}}^2 \cdot \epsilon/\hbar
c |\vec{\textsc{q}}_1|$.  The bandgaps predicted by the semiclassical
and quantum mechanical theories are thus similar for $\epsilon \sim
\hbar c |\vec{\textsc{q}}_1|$, which is also the regime where the
bandgaps are significant.  In the exactly-tuned case, the results are
identical, and one obtains
\begin{equation}
\Delta_{1d} = 2 \epsilon x_{01} \sqrt{\alpha \eta}\,.
\label{1D result}
\end{equation}

Actual 1\textsc{d}/2\textsc{d} lattices are more problematic since
each atomic excitation is coupled to photons with a continuum of
wavenumbers in the transverse direction, which smears out the gaps.
One might avoid this using an actual cavity in the transverse
direction, making the electromagnetic field effectively
1\textsc{d}/2\textsc{d}.

\section{Slow polariton modes}

The energy $E^0_{\textsc{q}}$ in (\ref{edge energies}) corresponds to
a polariton created by the operator $(a^\dagger_{\textsc{q}} -
a^\dagger_{\textsc{q}+\textsc{g}'}) / \sqrt{2}$.  This remains an
exact polariton state when we include higher BZs in the effective
Hamiltonian.  (In fact, there is a family of such states for each pair
of BZ boundaries.)  These ``purely photonic'' polaritons are
reminiscent of ``dark states'' in electromagnetically induced
transparency (EIT)\cite{slow light}, since (\ref{edge Hamiltonian})
is identical to the EIT effective Hamiltonian with the atomic
excitation and two photon modes acting as the levels of the $\Lambda$
system.  In EIT, a ``dark state'' arises: a coherent superposition of
atomic levels that does not couple to the radiation.  The analog in
our case is a non-interacting photonic state, with no atomic
component.  Its classical limit is a standing electromagnetic wave
commensurate with the lattice.  Since the laser light that supports
the lattice always falls exactly on the BZ boundary\cite{Deutsch},
the stability of the optical lattice relies on the existence of such
standing wave modes; other modes are Bragg reflected away.  In a
sense, the lattice ``selects'' the standing wave modes from the
incoming laser light.  Similar modes have been observed in other
resonant PhC systems\cite{MQW,MQW2,MQW3}.  We have shown here that in
the self-consistent limit of complete quantum coherence and low
excitation density, this selection takes place at the quantum state
level.  Only the purely photonic polaritons can support a macroscopic
population, since they are the only elementary excitations of the
interacting system with zero atomic component.

\begin{figure}
\includegraphics[width=0.48\textwidth]{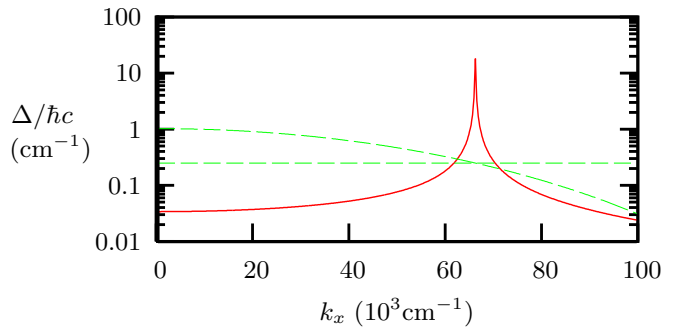}
\caption{(Color online) Photonic gap at wavevectors $\vec{\textsc{q}}
  = [k_x, \pi/\ell, 0]$ along the BZ boundary, for $\epsilon = 3$\,eV,
  $x_{01} = 2$\,\AA, and $\pi/\ell = 0.9\,\epsilon/\hbar c = 1.4
  \times 10^5\,\textrm{cm}^{-1}$\,(red-detuned).  The dashed lines
  show $k_y$ vs.~$k_x$ for the surface $|\vec{k}| = \epsilon/\hbar c$
  and the BZ boundary; here, the ordinate is not drawn to scale.  The
  gap is largest at the intersection of the two surfaces,
  i.e.~$|\vec{\textsc{q}}| = \epsilon/\hbar c$.}
\label{photonic gap figure}
\end{figure}

In the 3\textsc{D} system, there is a family of purely photonic
polaritons everywhere on the boundary of the first BZ.  Remarkably,
these states are attached to the slow, ``atomic'' branch of the
dispersion curve.  These appear to be analogs of the slow,
non-degenerate, longitudinal electromagnetic modes that appear in the
classical t-matrix calculation of Coevorden
\textit{et.~al.}\cite{Coevorden} \, Our model shows that the photonic
component along this branch goes continuously from nearly zero to
unity as we approach the BZ boundary, as shown in Fig. \ref{polariton
  figure}(c) and (d).  Therefore, by exciting polaritons over a range
\begin{equation}
  \left| \frac{\epsilon - \hbar c |\vec{\textsc{q}}|}{\hbar c} \right|
  \sim \frac{C_{\vec{\textsc{q}}}}{\hbar c}
  \sim 10\,\textrm{cm}^{-1}
\end{equation}
around wavevector $\vec{\textsc{q}}$, one could create a wavepacket
that propagates slowly but has low atomic excitation density.

\section{Conclusion}

We have presented a quantum model for an atomic lattice that applies
directly to optical lattices filled with cold atoms, containing
behavior similar to other resonant PhC systems.  The system possesses
two gaps (polaritonic and photonic) at each angle, and can be tuned so
that the gaps meet to create a combined gap orders of magnitude larger
than the individual detuned gaps, in a process analogous to
microcavity Rabi splitting; however, there does not exist a complete
gap.  The quantum analysis yields a branch of the dispersion curve
that has low group velocity and atomic component vanishing at the BZ
boundary.

These effects could be explored with alkali atoms held in a cubic
lattice made by near-IR light, by introducing a probe beam at an angle
to the axis of the lattice.  One should choose an atomic transition
$\epsilon$ such that $1\le \epsilon \ell / \pi \hbar c \le \sqrt{3}$,
where $\ell$ is the lattice period, and use probe wavevectors with
magnitude lying in a range $\Delta/\hbar c \sim 10\,\mbox{cm}^{-1}$
around $|\vec{q}| = \epsilon/\hbar c$, at an angle $\cos^{-1}(\pi
\hbar c/\epsilon \ell)$ to a lattice axis (Fig. \ref{photonic gap
  figure}).  Although the present theory applies to an infinite
lattice, the predicted frequency shifts may be observable close to the
atomic resonance, even in a lattice of about 100 atoms on a side.

We have treated the atomic positions as fixed, as would be the case
for a strongly-confining optical lattice where the rate at which each
atom tunnels to a different lattice site is negligible compared to the
radiative lifetime.  The presence of non-zero hopping amplitudes would
add an imaginary part to the polariton energies, proportional to the
tunneling rate.  The size of the band gaps would be reduced by the
corresponding amount.

Finally, it is interesting to note that the gap in Eq. (\ref{large
  gap}), which scales as $\epsilon$ relative to the photon energy, is
$O(10^{-2})\,\epsilon$ for X-rays.  Aspects of this theory might thus
be applicable to crystalline solids in the X-ray regime, where a
similar effect---superradiant scattering enhancement due to nuclear
resonances---is known to exist\cite{Smirnov}.

\begin{acknowledgments}
We would like to thank R.W.~Boyd, J.~D.~Joannopoulos, V.~Vuletic,
A.~Gordon, and F.~X.~Kaertner for helpful discussions.
\end{acknowledgments}

\end{document}